\documentclass[reprint,aps,prl,amsfonts,showpacs,letterpaper]{revtex4-1}

\usepackage{CJK}
\usepackage{graphicx, multirow,bm, amsmath}
\usepackage{hyperref}
\usepackage[usenames,dvipsnames]{color}

\begin{document}
\begin{CJK*}{UTF8}{bsmi}
	\title{Lattice anharmonicity and thermal conductivity from compressive sensing of first-principles calculations}

	\author{Fei Zhou(周非)}
	\affiliation{Physical and Life Sciences Directorate, Lawrence Livermore National Laboratory, Livermore, California 94550, USA}

	\author{Weston Nielson}
	\author{Yi Xia}
	\author{Vidvuds Ozoli\c{n}\v{s}}
	\affiliation{Department of Materials Science and Engineering, University of California, Los Angeles, California 90095-1595, USA}

	\date{\today}

\begin{abstract}
First-principles prediction of lattice thermal conductivity $\kappa_L$ of strongly anharmonic crystals is a long-standing challenge in solid state physics. Making use of recent advances in information science, we propose a systematic and rigorous approach to this problem, compressive sensing lattice dynamics (CSLD). Compressive sensing is used to select the physically important terms in the lattice dynamics model and determine their values in one shot. Non-intuitively, high accuracy is achieved when the model is trained on first-principles forces in {\it quasi-random\/} atomic configurations. The method is demonstrated for Si, NaCl, and Cu$_{12}$Sb$_4$S$_{13}$, an earth-abundant thermoelectric with strong phonon-phonon interactions that limit the room-temperature $\kappa_L$ to values near the amorphous limit.
\end{abstract}
\maketitle
\end{CJK*}

To a large extent, thermal properties of crystalline solids are determined by the vibrations of their constituent atoms. Hence, an accurate description of lattice dynamics is essential for fundamental understanding of the structure, thermodynamics, phase stability and thermal transport properties of solids. The seminal work of Born and Huang \cite{Born-Huang1954Dynamical} forms the theoretical basis of our understanding of harmonic vibrations and their relation to elastic properties. With the advent of efficient density-functional theory (DFT) based methods for solving the Schr\"odinger's equation, several {\it ab initio\/} methods for studying {\it harmonic} phonon properties of solids have been proposed, such as the frozen phonon approach \cite{Wendel1978,Ho1984}, supercell small displacement method \cite{Kunc1982,Parlinski1997} and the density-functional perturbation theory (DFPT) \cite{Baroni2001}. Due to these developments, {\it ab initio\/} calculations of the harmonic phonon dispersion curves and phonon mode Gr\"uneisen parameters have become routine.

A systematic approach to {\it anharmonicity} has been more difficult to develop. Anharmonic effects are key to explaining phenomena where phonon-phonon collisions become important, such as in the study of lattice thermal conductivity, $\kappa_L$, a key quantity for optimizing the performance of electronic materials, thermal coatings, and thermoelectrics \cite{Zebarjadi2012}.
For weakly anharmonic systems, the interaction processes involving three phonons are dominant, such as decay of a phonon into two lower-energy phonons or combining two phonons to create a higher-energy phonon. Their effects on phonon frequencies and lifetimes can be evaluated using the first-order perturbation theory (PT) \cite{Horton1974dynamical, Maradudin1962}, and $\kappa_L$ can then be obtained by either using the relaxation time approximation or solving the Boltzmann transport equation \cite{Omini1996}. The computational feasibility and physical accuracy of these methods are well established \cite{Debernardi1995,Broido2007,Garg2011,Esfarjani2011}. Unfortunately, PT tends to be computationally expensive for solids with large, complex unit cells, and its ability to handle strong anharmonicity is insufficient, especially when the harmonic phonon dispersion contains imaginary frequencies or when phonon scattering becomes so intense that $\kappa_L$ saturates at its theoretical minimum \cite{Cahill1992}. Many interesting and technologically relevant materials belong to this class, e.g. ferroelectrics and thermoelectrics with ultra-low $\kappa_L$ \cite{Nielsen2013}. In these cases, a general and efficient non-perturbative approach that can accurately describe four-phonon and higher-order interactions, is needed. The ``$2n+1$'' theorem of DFPT \cite{Gonze1989} can be used to calculate the 4th- and higher-order terms, but the computations are cumbersome and require specialized codes which, to the best of our knowledge, are not available for $n >1$.

In this paper, we introduce an approach to building lattice dynamical models which can treat compounds with large, complex unit cells and strong anharmonicity, including those with harmonically unstable phonon modes. Our approach, compressive sensing lattice dynamics (CSLD), determines anharmonic force constants from standard DFT total energy calculations. We utilize compressive sensing (CS), a technique recently developed in the field of information science for recovering sparse solutions from incomplete data \cite{Candes2008introduction}, to determine which anharmonic terms are important and find their values simultaneously. A non-intuitive prescription based on CS to generate DFT training data is given. We show that CSLD is efficient, general and robust through a few prototypical case studies.

The starting point is a Taylor expansion of the total energy in powers of atomic displacements,
\begin{eqnarray}
\label{eq:Taylor}
V &=& V_{0} + \Phi_{\mathbf{a}} u_{\mathbf{a}} + \frac{\Phi_{\mathbf{ab}} }{2} u_{\mathbf{a}} u_{\mathbf{b}} + \frac{ \Phi_{\mathbf{abc}}}{3!}  u_{\mathbf{a}} u_{\mathbf{b}}u_{\mathbf{c}}+ \cdots \label{eq:potentialexpansion}
\end{eqnarray}
where $u_{\mathbf{a}} \equiv u_{a,i}$ is the displacement of atom $a$ at a lattice site ${\bf R}_a$ in the Cartesian direction $i$, the 2nd-order expansion coefficients $\Phi_{\mathbf{ab}} \equiv \Phi_{ij}(ab) = \partial^{2} V/\partial u_{\mathbf{a}} \partial u_{\mathbf{b}}$ determine the phonon dispersion in the harmonic approximation, and $\Phi_{\mathbf{abc}} \equiv \Phi_{ijk}(abc) =  \partial^{3} V/\partial u_{\mathbf{a}} \partial u_{\mathbf{b}} \partial u_{\mathbf{c}}$, etc., are third- and higher-order anharmonic force constant tensors (FCTs). The linear term with $\Phi_{\mathbf{a}}$ is absent if the reference lattice sites represent mechanical equilibrium, and the Einstein summation convention over repeated indices is used throughout the paper.

Systematic fitting or direct calculation of the higher-order anharmonic terms in Eq.~(\ref{eq:Taylor}) is challenging due to combinatorial explosion in the number of tensors $\Phi(a_{1} \cdots a_{n})$ with increasing order $n$ and maximum distance between the sites $\{a_{1}, \ldots, a_{n}\}$. Since it is not {\it a priori\/} obvious where to truncate this expansion, one needs to rely on physical intuition, which can only be gained on a case-by-case basis through time-consuming cycles of model construction and cross-validation. As a result,  anharmonic FCTs have been calculated only for relatively simple crystals and weak anharmonicity \cite{Esfarjani2008,Esfarjani2011,Hellman2013}.

We have recently shown that a similar problem in alloy theory, the cluster expansion (CE) method for configurational energetics \cite{Sanchez1984PA334, deFontaine199433}, can be solved efficiently and accurately using compressive sensing \cite{Nelson2013a, Nelson2013b}. CS has revolutionized information science by providing a mathematically rigorous recipe for reconstructing $S$-sparse models (i.e., models with $S$ nonzero coefficients out of a large pool of possibles, $N$, when $S \ll N$) from a set of only $O(S)$ data points \cite{Candes2005,Candes2006a,Candes2006b}. Given training data, CS automatically picks out the relevant expansion coefficients and determines their values {\it in one shot\/} by applying a mathematical technique which in essence is the Occam's razor for model choice. To see how this applies to lattice dynamics, we write down the force-displacement relationship for Eq.~(\ref{eq:Taylor}):
\begin{eqnarray}
\label{eq:force}
F_{\mathbf{a}}  = 
 - \Phi_{\mathbf{a}}  - \Phi_{\mathbf{a}\mathbf{b}} u_{\mathbf{b}} -  \Phi_{\mathbf{a}\mathbf{b}\mathbf{c}}  u_{\mathbf{b}}u_{\mathbf{c}}/2 - \cdots.
\end{eqnarray}
The forces can be obtained from first-principles calculations using any general-purpose DFT code for a set of $L$ atomic configurations in a supercell. This establishes a linear problem $\mathbf{F} =  \mathbb{A} \mathbf{ \Phi}$ for the unknown FCTs, where
\begin{eqnarray}
\label{eq:FeqAPhi}
 \mathbb{A} = \begin{bmatrix} -1 & -u_{\mathbf{b}}^1  & - \frac{1}{2}  u_{\mathbf{b}}^1 u_{\mathbf{c}}^1 & \cdots \\
                                                   & \cdots & & \\
                                                -1 & -u_{\mathbf{b}}^L  & - \frac{1}{2}  u_{\mathbf{b}}^L u_{\mathbf{c}}^L & \cdots
                                             \end{bmatrix}
\end{eqnarray}
will be referred to as the sensing matrix. Its elements are products of atomic displacements corresponding to distinct terms in the force expansion Eq.~\eqref{eq:force}; different training configurations are labeled by superscript $u_{\mathbf{b}}^i$. Each row corresponds to a calculated force component on one of the atoms, and the total number of rows is $M=3LN_\text{at}$, where $N_\text{at}$ is the number of atoms in the supercell. Columns corresponds to $N$ FCT components, which are arranged in a vector $\mathbf{\Phi}$. In practice, $N$ can far exceed $M$, which makes the linear problem Eq.~(\ref{eq:FeqAPhi}) underdetermined. A reasonable approach would be to choose $\mathbf{\Phi}$ so that it reproduces the training data $\mathbf{F}$ to a given accuracy with the smallest number of nonzero FCT components, i.e., by minimizing the so-called $\ell_0$ norm of the solution. Unfortunately, this is an intractable (``NP-hard'') discrete optimization problem.

CS solves the underdetermined linear problem in Eq.~(\ref{eq:FeqAPhi}) by minimizing the $\ell_1$ norm of the coefficients, $\| \mathbf{\Phi} \|_1 \equiv \sum_{I} | \Phi_{I} |$, while requiring a certain level of accuracy for reproducing the data. The $\ell_1$ norm serves as an approximation to the $\ell_0$ norm and results in a computationally tractable convex optimization problem.  Mathematically, the solution is found as
\begin{eqnarray}
\label{eq:CSfit}
\mathbf{\Phi}^\text{CS} &=& {\arg \min}_\mathbf{\Phi} \| \mathbf{\Phi} \|_1 + \frac{\mu}{2} \| \mathbf{F} -  \mathbb{A} \mathbf{ \Phi} \|^2_2  \nonumber \\
&=&  {\arg \min}_\mathbf{\Phi} \sum_{I} | \Phi_{I} | + \frac{\mu}{2} \sum_{ai} \left( F_{ai} - A_{ai, J} \Phi_J \right)^2,
\end{eqnarray}
where the second term is the usual sum-of-squares Euclidian $\ell_2$ norm of the fitting error for the training data (in this case, DFT forces). The $\ell_1$ term drives the model towards solutions with a small number of nonzero elements, and the parameter $\mu$ is used to adjust the relative weights of the $\ell_1$ and $\ell_2$ terms (see below). 
CSLD has several advantages over other methods for building models of lattice dynamics: it does not require prior physical intuition to pick out potentially relevant FCTs, the fitting procedure is very robust with respect to both random and systematic noise \cite{Candes2006a}, and it gives an efficient prescription for generating training data.

A full account of the technical details of our approach will be given in a separate publication, and here we only describe the key features. Higher values of $\mu$ will produce a  least-squares like fitting at the expense of denser FCTs that are prone to over-fitting, while small $\mu$ will produce very sparse under-fitted FCTs, degrading the quality of the fit. The optimal $\mu$ value that produces a model with the highest predictive accuracy lies in-between and can be determined by monitoring the predictive error for a leave-out subset of the training data not used in fitting \cite{Nelson2013a}. The predictive accuracy of the resulting model is then validated on a third, distinct set of DFT data, which we refer to as the ``prediction set''. Space group symmetry and translational invariance conditions are used to reduce the number of independent FCT elements \cite{Horton1974dynamical}; the latter are also important for momentum conservation according to the Noether's theorem. These linear constraints are applied algebraically by constructing a null-space matrix.
For polar insulators, the long-range Coulomb interactions can be treated separately by calculating the Born effective charges and dielectric tensors. The long-range contributions are then subtracted from $\mathbf{F}$, ensuring that the remaining FCTs are short-ranged \cite{Gonze1997PRB10355}.

A key ingredient of CSLD is the choice of atomic configurations for the training and prediction sets. It is intuitively appealing to use snapshots from {\it ab initio} molecular dynamics (AIMD) trajectories since they represent physically relevant low-energy configurations. However, these configurations give rise to strong cross-correlations between the columns of $\mathbb{A}$ (i.e., high mutual coherence of the sensing matrix \cite{Donoho2001ITIT2845}), which decreases the efficiency of CS due to the difficulty of separating correlated contributions to $\mathbf{F}$ from different FCTs. One of the most profound results of CS is that a near-optimal signal recovery can be realized by using sensing matrices $\mathbb{A}$ with {\it random\/} entries that are independent and identically distributed (i.i.d.) \cite{Candes2008introduction}. In this case, the contributions from different FCTs are uncorrelated and can be efficiently separated using Eq.~(\ref{eq:CSfit}). For the discrete orthogonal basis in the CS cluster expansion \cite{Nelson2013a,Nelson2013b}, i.i.d. sensing matrices $\mathbb{A}$ could be obtained by enumerating all ordered structures up to a certain size and choosing those with correlations that map most closely onto quasi-random vectors. This strategy is difficult to adapt for CSLD since the Taylor expansion employs non-orthogonal and unnormalized basis functions of a continuous variable, $u^{n}$. To solve this conundrum, we combine the physical relevance of MD trajectories with the mathematical efficiency of CS by adding random displacements ($\sim$ 0.1 \AA) to each atom in
well-spaced MD snapshots.
To guarantees that all the terms in the $\ell_1$ norm have the same unit of force, Eq.~(\ref{eq:force}) is scaled by $\Phi \rightarrow \Phi u_{0}^{n-1}$ and $u \rightarrow u/u_{0}$, where $n$ is the order of the FCT and $u_{0}$ is a ``maximum'' displacement chosen to be on the order of the amplitude of thermal vibrations. This procedure was found to significantly decrease cross-correlations between the columns of $\mathbb{A}$ and resulted in stable fits.

\begin{figure}[t]
\includegraphics[width = 0.75 \linewidth]{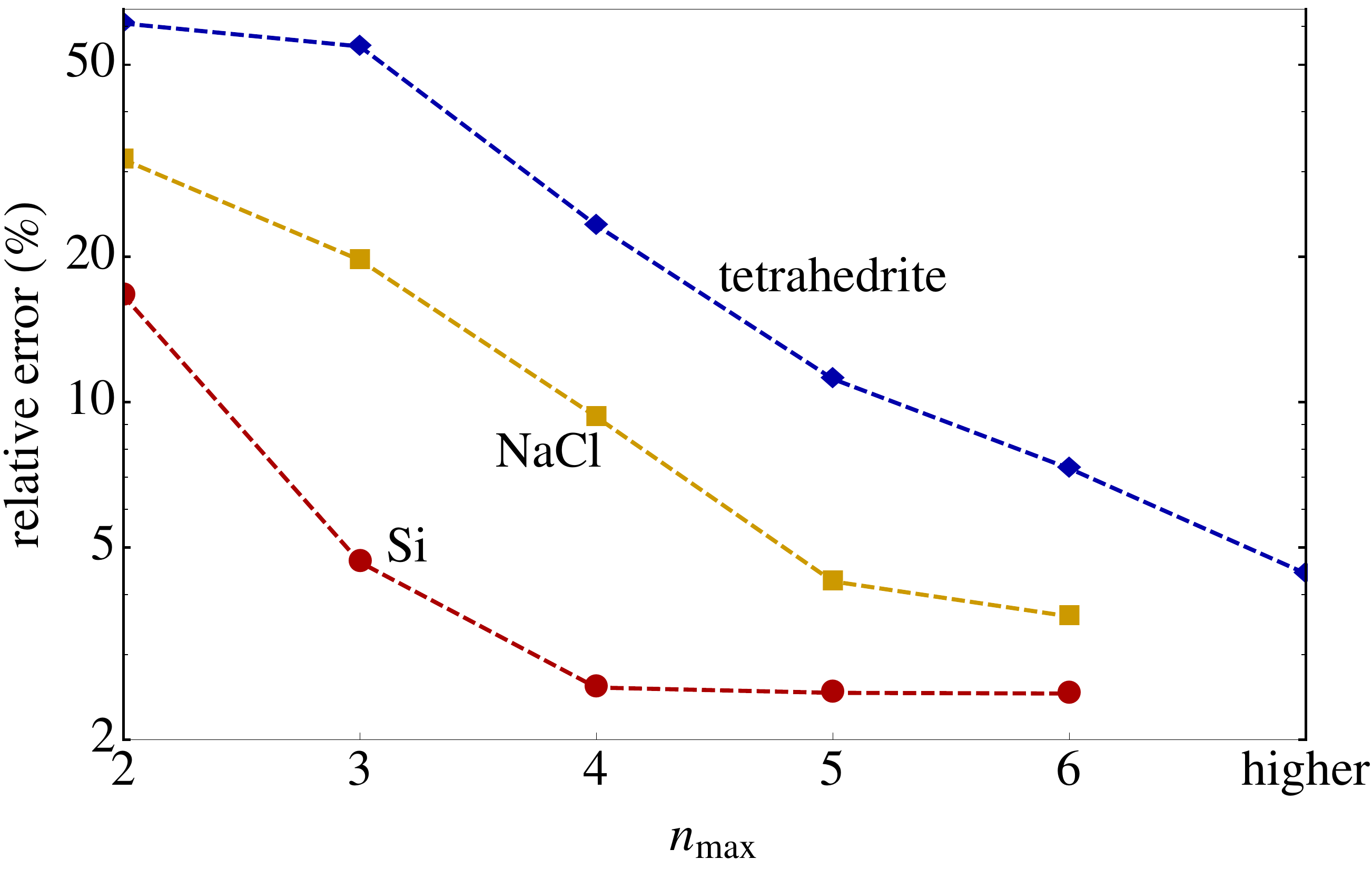}
\caption{Relative CSLD prediction error of force components from AIMD snapshots at 300 K not involved in fitting. For each system, a fitting of up to 6th order was performed, and errors using FCTs up to $n_{\max} (= 2, \dots)$ are shown.  Higher order expansions are used in tetrahedrite \cite{Note2}.}
\label{fig:err-vs-nmax}
\end{figure}
\begin{figure}[htp]
    \includegraphics[width =0.92 \linewidth]{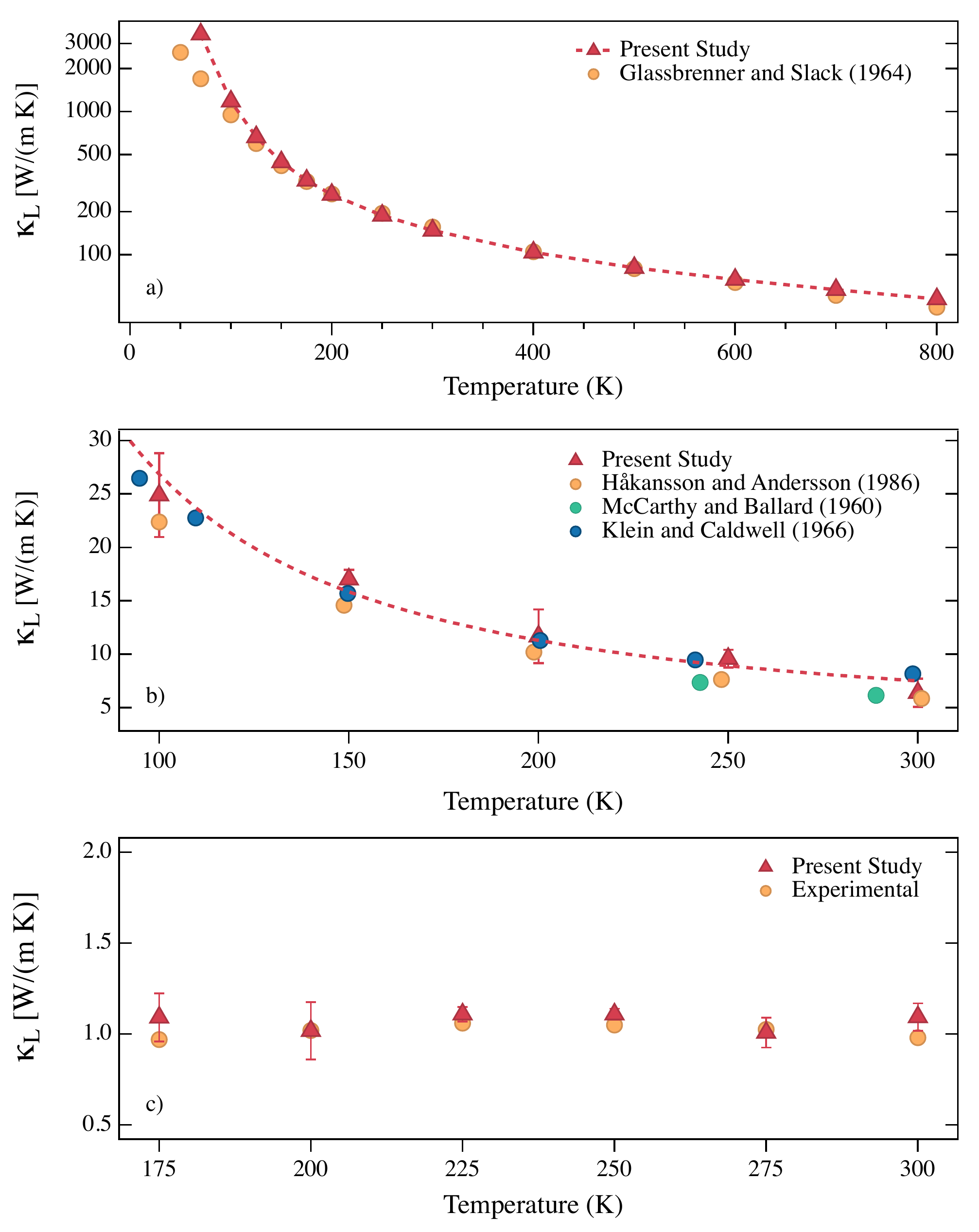}
    \caption{Temperature dependence of calculated $\kappa_L$ (triangles) for (a) Si from Boltzman transport equations using 3$\text{rd}$ order FCTs; (b) NaCl and (c) Cu$_{12}$Sb$_4$S$_{13}$ by MD simulation using up to 6$\text{th}$-order FCTs. The dashed line is an inverse power fit of the calculation, and circles are experimental data from Ref.~\onlinecite{Glassbrenner1964PRA1058}, Refs.~\onlinecite{Hakansson1986355,Klein:1966qf,McCarthy:1960qv} and Ref.~\onlinecite{Lu2012AEM342}, respectively. }
    \label{fig:all-TC}
\end{figure}

We begin by demonstrating the accuracy of our approach for two relatively simple cases, Si and NaCl. DFT calculations were performed using the Perdew-Becke-Ernzerhof (PBE) functional \cite{Perdew1996PRL3865} and projector-augmented wave (PAW) potentials \cite{Blochl1994PRB17953} as implemented in the VASP code \cite{Kresse1999PRB1758}. An overview of the predictive accuracy of CSLD is shown in Fig.~\ref{fig:err-vs-nmax}. We included up to the 6-th order FCTs for a total of 712 (Si) and 1375 (NaCl) symmetrically distinct elements. CS using Eq.~(\ref{eq:CSfit}) found 258 and 199 non-zero FCT elements, respectively. The errors decrease when higher order FCT parameters are considered. Anharmonic terms account for an increasing amount of the improved accuracy in Si, NaCl and tetrahedrite (to be discussed later), reflecting increasing anharmonicity. Phonon dispersion curves (Supplemental Material) using the CSLD pair force constants are in excellent agreement with experiment, validating our method on the harmonic level. We then used first-order PT \cite{Horton1974dynamical, Maradudin1962} to calculate phonon lifetimes of Si (Supplemental Material), which are in excellent agreement with other first-principles PT based studies \cite{Esfarjani2011, Hellman2013}. Lattice thermal conductivity $\kappa_L$ of Si (Fig.~\ref{fig:all-TC}a) was obtained with the ShengBTE code \cite{ShengBTE2014} and found to be in good agreement with experimental data \cite{Glassbrenner1964PRA1058},  validating the numerical accuracy of our third-order FCTs.

 To test the performance of CSLD in calculating $\kappa_L$ of strongly anharmonic solids, a custom lattice molecular dynamics (LMD) program was developed with Eq.~(\ref{eq:potentialexpansion}) as the potential.  Multiple methods were implemented for calculating $\kappa_L$, including the Green-Kubo linear response formula \cite{Green:1954kx,Kubo:1957nr}, reverse non-equilibrium MD (RNEMD) \cite{Muller-Plathe:2004fr} and homogenous non-equlibrium MD (HNEMD) proposed by Evans \cite{Evans:1982qv}. While all methods yielded similar results, we found after extensive testing that HNEMD was the most efficient. In HNEMD, the equations of motion are modified so that the force on atom $a$ is given by
\begin{equation}
    \label{eq:eq-motion}
    \mathbf{F}_a= F_a - \sum\limits_{b} \mathbf{F}_{ab} \left( \mathbf{r}_{ab} \cdot \mathbf{F}_e \right) + \frac{1}{N} \sum\limits_{b,c} \mathbf{F}_{bc}  \left( \mathbf{r}_{bc} \cdot \mathbf{F}_e \right),
\end{equation}
where $F_a$ is the unmodified force calculated from Eq.~(\ref{eq:force}) and $\mathbf{F}_{ab}$ is the force on atom $a$ due to  $b$ \footnote{Contributions from third- and higher-order interactions to $\mathbf{F}_{ab}$ were obtained by partitioning the energy evenly among all interacting atoms, including repeated sites.}. The external field $ \mathbf{F}_e$ has the effect of driving higher energy (hotter) particles with the field and lower energy (colder) particles against the field, while a Gaussian thermostat is used to remove the heat generated by $ \mathbf{F}_e$. Using linear response, the average heat flux is given by
\begin{equation}
  \left<\mathbf{J}(t)\right> =  -\beta V \int\limits_0^t ds \left< \mathbf{J}(t-s)  \otimes  \mathbf{J}(0)  \right> \cdot \mathbf{F}_e.
\end{equation}
As $\mathbf{F}_e \rightarrow 0$ and $t \rightarrow \infty$, one recovers the Green-Kubo formula \cite{Green:1954kx,Kubo:1957nr}. For cubic systems the external field can be set to $\mathbf{F}_e  = (0,0,F_z)$, and we get
\begin{equation}
  \kappa_L = \frac{V}{k_B T^2} \int\limits_0^\infty dt \left< J_z(t) J_z(0) \right> = \lim\limits_{F_z \rightarrow 0} \frac{- \left< J_z(\infty) \right>}{TF_z}.
\end{equation}
The process then involves a series of simulations at varying external fields $\mathbf{F}_e$ and constant ${T}$, with a simple linear extrapolation to zero field resulting in the true $\kappa_L$.

Simulations were performed for NaCl between 100 and 300~K, with system sizes ranging from 512 to 4096 atoms. The lengths of the simulations ranged from 100 ps to 1 ns and all used a timestep of 1 fs.  At least four different values for $F_{z}$ were taken at a given $T$. The results obtained  are shown for NaCl in Fig.~\ref{fig:all-TC}(b).  Very good agreement is seen between the calculated and experimental values across the entire temperature range tested.

Finally, CSLD is applied to study anharmonic phonon dynamics and $\kappa_L$ in Cu$_{12}$Sb$_{4}$S$_{13}$,  a parent compound for the earth-abundant natural mineral tetrahedrite, which was recently shown to be a high-performance thermoelectric \cite{Lu2012AEM342}.  One of its key advantages is an exceptionally low $\kappa_L$, experimentally found to be $ \lesssim $ 1 W/(m K) in phase and compositionally pure samples \cite{Lu2012AEM342}. Furthermore, our previous calculation found several harmonically unstable phonon modes, pointing to very strong anharmonicity \cite{Lu2012AEM342}.  Cu$_{12}$Sb$_{4}$S$_{13}$ has a body-centered cubic (space group  $I\bar{4}3m$) structure with 29 atoms in the primitive cell, a large number that complicates the computation of FCTs using existing methods. For example, there are 188 distinct atomic pairs  within a radius of $a=10.4$~{\AA}, 116 triplets within $a/2$, etc. Taking into account the $3^{n}$ elements of each tensor, the number of unknown coefficients is very large (55584 in our setting including up to 6$\text{th}$-order terms). After symmetrization, this is reduced to $N=3188$, which still represents a formidable numerical challenge.

\begin{figure}[htp]
\includegraphics[width =0.39 \linewidth]{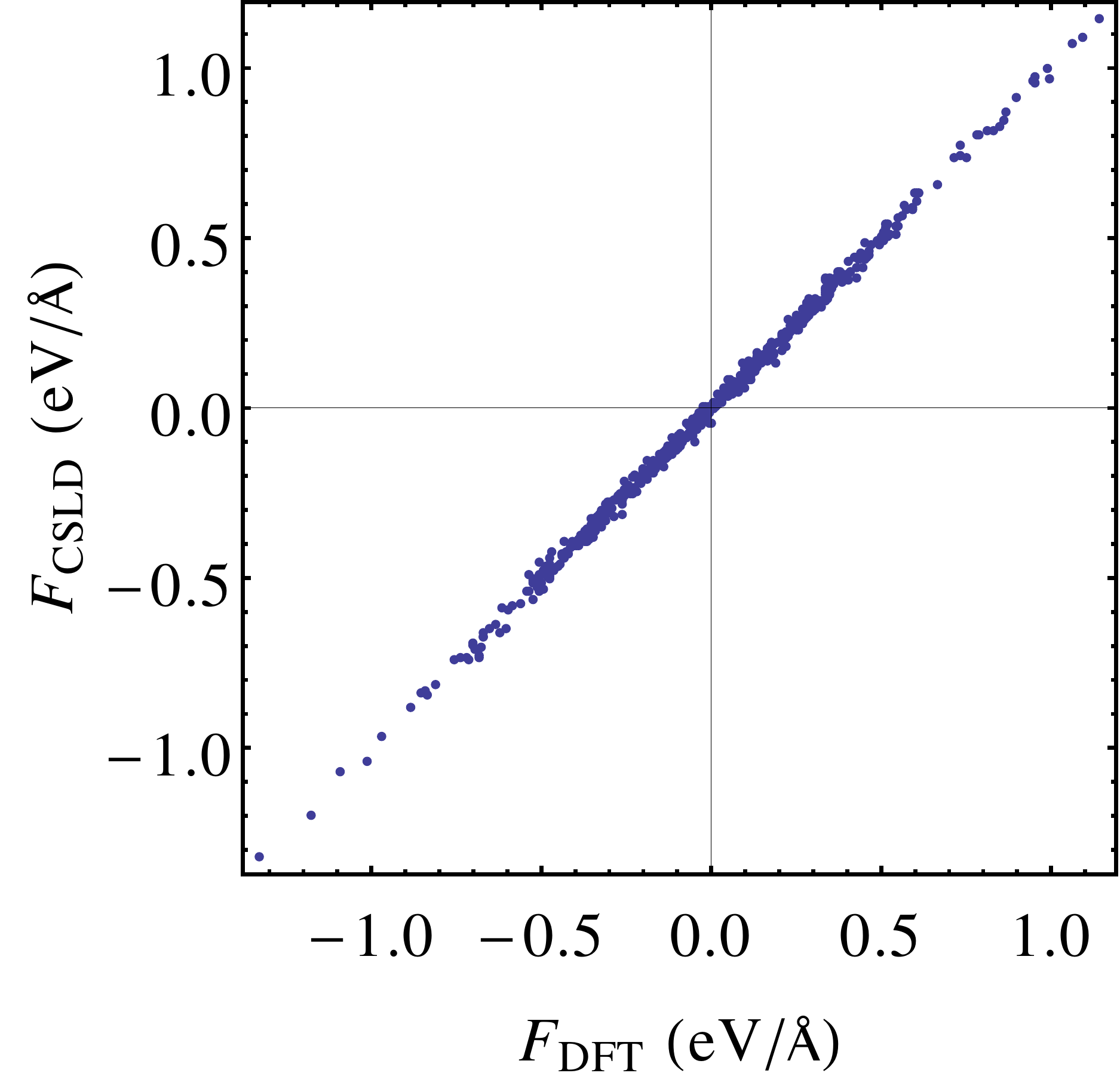}
\includegraphics[width =0.59 \linewidth]{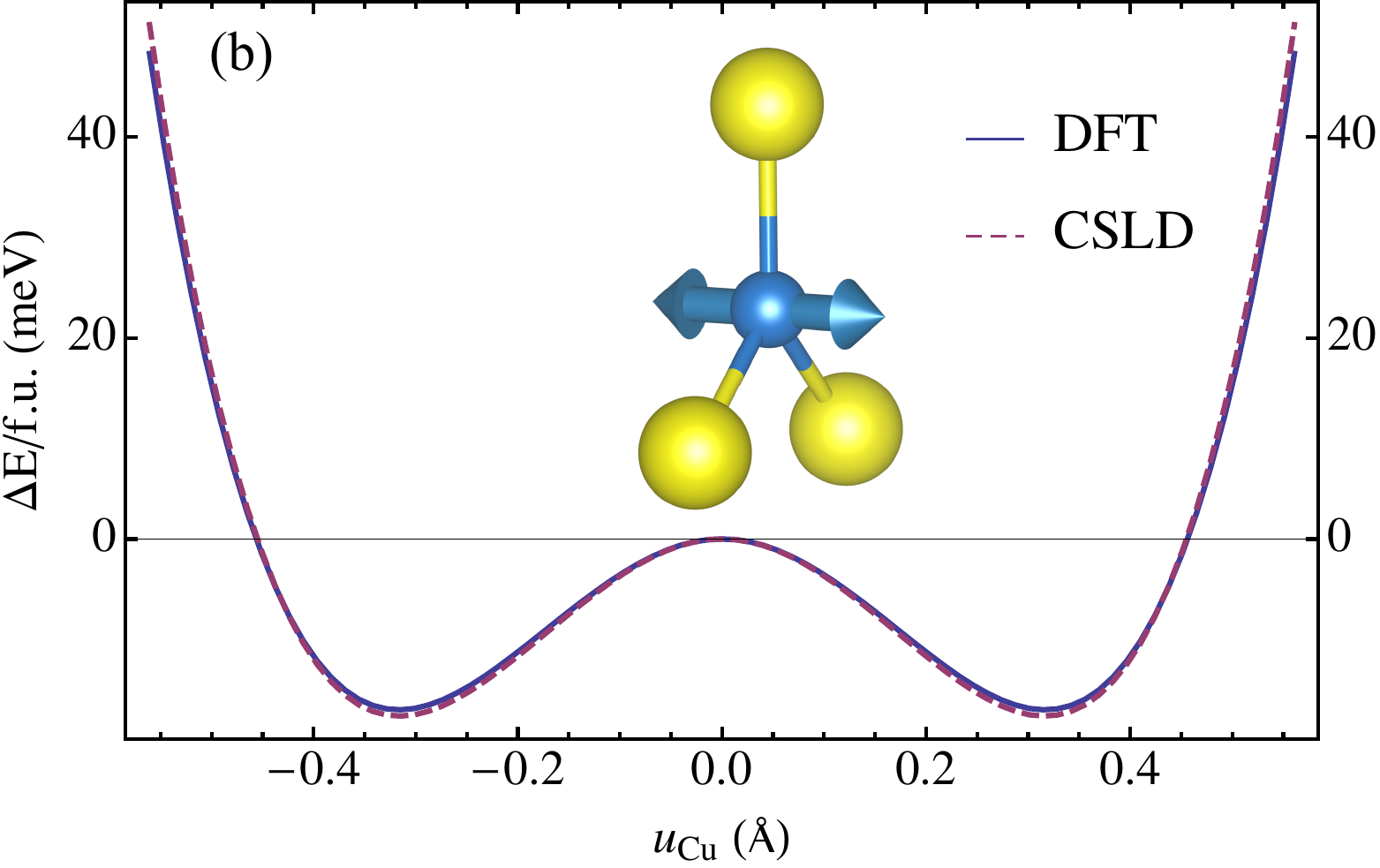}
\caption{Comparison of CSLD predictions with DFT data for tetrahedrite: (a) force at 300 K, (b) relative energy per formula unit of an unstable optical mode involving out-of-plane displacements of trigonally coordinated copper atoms (blue) bonded to sulfur (yellow sphere). DFT and CSLD are shown as solid and dashed lines, respectively.}
\label{fig:force-prediction}
\end{figure}

High order interactions were added \footnote{A series expansion in orthogonal polynomials for interactions of bonded atoms are determined simultaneously with the FCTs (details in a forthcoming paper)} to ensure highly accurate forces (see also Fig.~\ref{fig:err-vs-nmax}).  A  large number of non-zero FCT elements (2101) were obtained by CSLD.   Fig.~\ref{fig:force-prediction}a shows the overall accuracy of the model over a prediction set from {\it ab initio} MD snapshots at 300 K. The root-mean-square error of the predicted force components is  0.02 eV/{\AA}, or 4\%.
The phonon dispersion calculated from the pair FCTs features unstable modes and is in good agreement with our previous DFPT calculations (see Supplemental material) \cite{Lu2012AEM342}, again validating CSLD at the harmonic level. Figure~\ref{fig:force-prediction}b
shows the DFT potential energy surface (solid line) along an unstable $\Gamma$ point mode involving displacement of trigonally coordinated Cu atoms (inset). The double-well behavior  points to strong 4th-order anharmonicity. Our CSLD model (dashed line) is able to reproduce the potential energy to an absolute accuracy of 2 meV.

The HNEMD method was used to calculate $\kappa_L$ of Cu$_{12}$Sb$_{4}$S$_{13}$, employing the same approach as described for NaCl above. All simulations were done with a supercell of 464 atoms and a minimum of 4 separate external fields at each temperature.  The HNEMD results are compared with the experimental $\kappa$ from Lu {\it et al.\/} \cite{Lu2012AEM342} with electronic contributions subtracted in Fig.~\ref{fig:all-TC}(c).  Once again, very good agreement is seen across the entire temperature range tested.
This example shows that CSLD extends the accuracy of DFT to treat lattice dynamics of compounds with large, complex unit cells and strong anharmonic effects previously beyond the reach of non-empirical studies.

In conclusion, CSLD is a powerful tool for highly anharmonic lattice dynamics in complex materials based on the robust and mathematically rigorous framework of compressive sensing and compressive sampling. The main advantage of CSLD over the current methods is that it is widely applicable, computationally efficient, systematically improvable and straight-forward to implement. Importantly, it works with general-purpose DFT codes and can be used in an automated manner, with minimal human intervention.
This technical development is a big step towards systematic, automated calculations of thermal transport properties for a wide variety of crystalline compounds, enabling computational design and discovery of new high-performance materials. Beyond lattice thermal conductivity, we expect CSLD to be useful in a wide range applications where strong anharmonicity plays a key role, such as ferroelectric phase transitions and temperature induced structural phase transformations, including martensitic transformations.

\begin{acknowledgments}
The authors gratefully acknowledge discussions with B.\ Sadigh and financial support for general method development from the National Science Foundation under Award No.\ DMR-1106024. CSLD and LMD code development and calculations related to Cu$_{12}$Sb$_{4}$S$_{13}$ were performed as part of the Center for Revolutionary Materials for Solid State Energy Conversion, an Energy Frontier Research Center funded by the US DOE, Office of Science, Basic Energy Sciences under Award No.\ DE-SC0001054. We are grateful to X.~Lu and D.~T.~Morelli for providing low-temperature $\kappa_L$ data for tetrahedrite. Part of the work by F.Z. was performed under the auspices of the US DOE by Lawrence Livermore National Laboratory under Contract DE-AC52-07NA27344. We used computing resources at NERSC, which is supported by the US DOE under Contract No. DE-AC02-05CH11231. 
\end{acknowledgments}

%

\newpage
\clearpage
\onecolumngrid
\appendix
\renewcommand\thefigure{S\arabic{figure}}
\setcounter{page}{1}
\setcounter{figure}{0}
\renewcommand{\thesubsection}{\Alph{subsection}}

\section{Supplemental Material}
\setcounter{secnumdepth}{2}
\subsection{The procedure of model building} \label{sec:app:CSLD}
Our CSLD method for lattice anharmonicity and thermal conductivity calculations follow the procedure as outlined in Fig.~\ref{fig:flow}. Note that in this procedure:
\begin{figure}[htp]
\includegraphics[width =0.33 \linewidth]{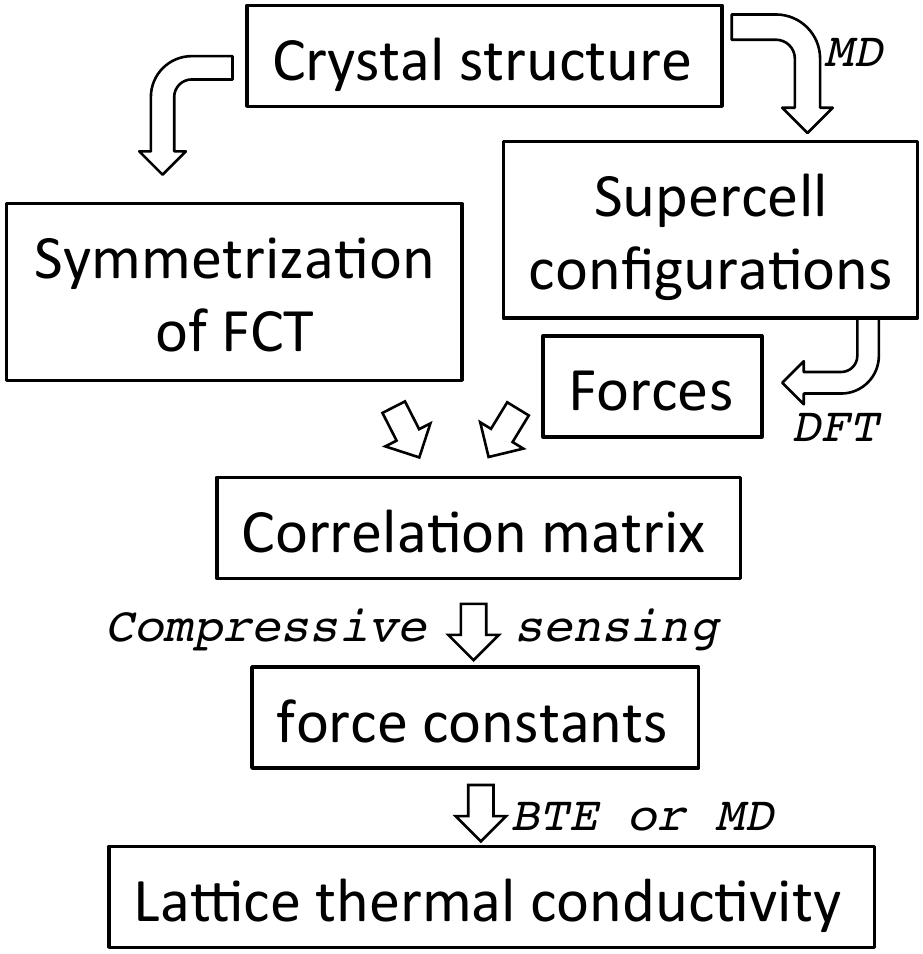}
\caption{Flow chart of calculating force constants and lattice thermal conductivity.}
\label{fig:flow}
\end{figure}
\begin{itemize}
\item The supercell structures used for training the lattice dynamical models are obtained as follows:
\begin{enumerate}
\item {\it Ab initio} or classical molecular dynamics at high temperature is performed to obtain supercell snapshots. In this work we run Born-Oppenheimer {\it ab initio} MD at around 600-800 K for the case studies. Relatively small cutoff energy and few K-points may be used in these MD simulations since we are primarily interested in training structures, not accurate energy and forces. In each material,  5--15 snapshots were taken with time intervals of $100 \sim 150 $ fs  from the MD trajectory.
\item Random small displacements ($\lesssim 0.1$ \AA) are given to each atom in the snapshot. These displacements constitute the training structure.
\end{enumerate}
\item Similar to the supercell small displacement method for phonons, high quality DFT calculations are performed for each snapshot to obtain atomic forces. Care was taken to ensure they are well converged with respect to K-point sampling, plane-wave cutoff energy, etc.
\item The correlation matrix is computed from the structure-force relationship taking into account the symmetry properties and constraints of the force constants, e.g. space group symmetry and translational invariance.
\item The pre-conditioned split-Bregman compressive sensing algorithm was used to fit the very large number of parameters and enhance the numerical stability of fitting force constants. The fitting accuracy is validated with a separate set of ``hold-out'' structures that are not included in the training set. We monitor the prediction accuracy over a range of $\mu$ values to determine the optimal $\mu$.
\item The obtained FCT's are also applied to predict a third, independent set of structures/force components to check the accuracy.
\end{itemize}

\subsection{Simple example of the compressive sensing approach}
\begin{figure}[htp]
\includegraphics[width =0.63 \linewidth]{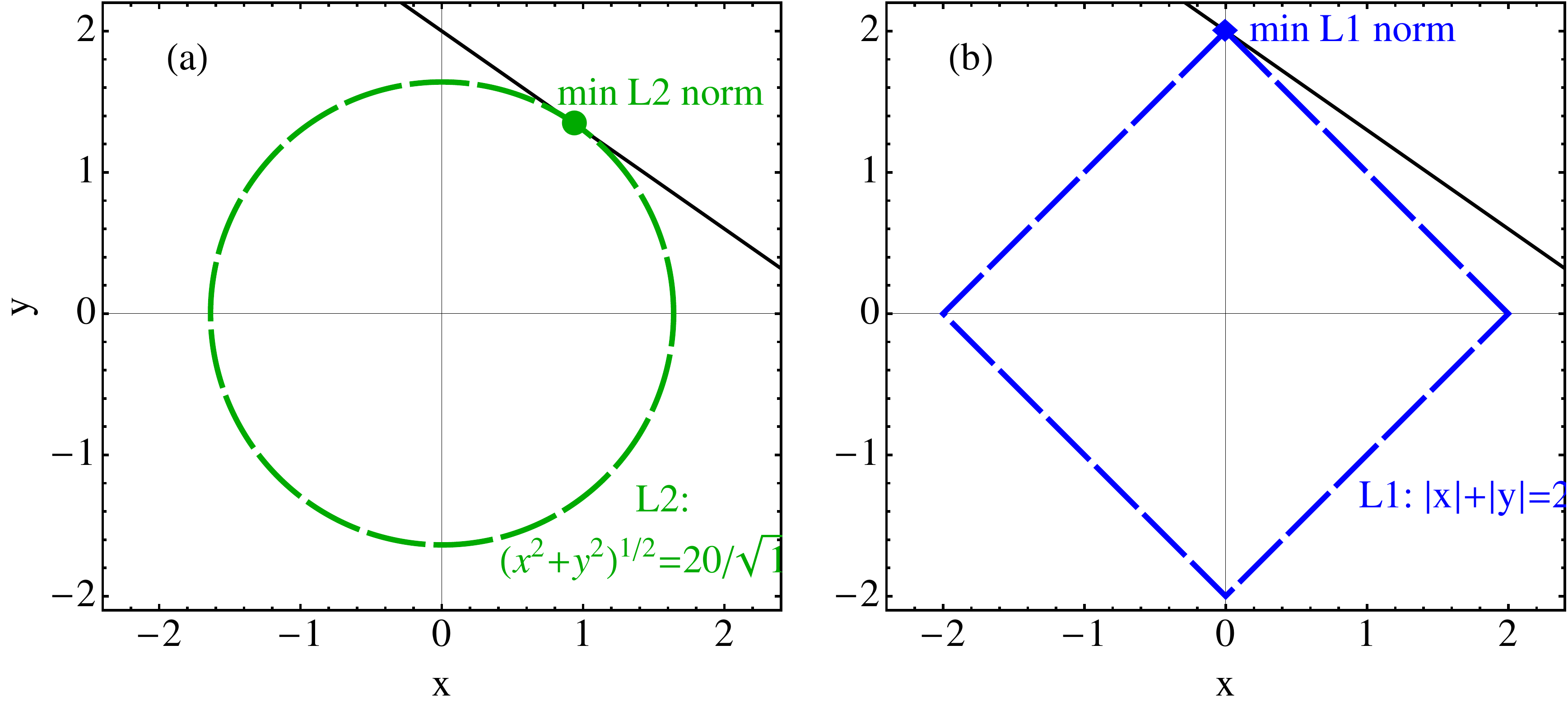}
\caption{An example to comparise of the iso-surface of (a) the $\ell_2$ norm (Euclidian norm $\sqrt{x^{2}+y^{2}}$) and (b) $\ell_1$ norm  ($|x|+|y|$) in 2D, shown as dashed lines. The under-determined linear equation is $7x+10y=20$, designated by the solid straight line. The filled point  where the iso-surface touches the straight line is the solution with the smallest $\ell_2$ or $\ell_1$ norm.}
\label{fig:L1L2}
\end{figure}

We use the compressive sensing approach to treat the under-determined linear problem $\mathbf{f} =  \mathbb{A} \mathbf{ u}$ while minimizing the $\ell_1$ norm $\| \mathbf{u} \|_1 \equiv \sum_{i} | u_{i} |$.
At the very core of the compressive sensing approach, one makes the assumption that the solution vector is sparse, or has few nonzero components. The $\ell_1$ norm is an effective constraint to direct the  search for optimal fitting towards the most sparse solution. To illustrate how compressive sensing finds an optimal sparse solution, examine the trivially simple, underdetermined system of $7x+10y=20$, as shown in Fig.~\ref{fig:L1L2}. This example was previously used to demonstrate the Bayesian compressive sensing approach for cluster expansion [L. J. Nelson, V. Ozolins, C. S. Reese, F. Zhou, and G. L. W. Hart, Phys. Rev. B 88, 155105 (2013)]. Minimization of the $\ell_2$ norm by staying on the straight line corresponds to a dense solution: both $x$ and $y$ are non-vanishing (filled circle in Fig.~\ref{fig:L1L2}a). Minimizing the $\ell_1$ norm $|x|+|y|$ corresponds to an optimal sparse solution $\{x=0, y=2\}$ (filled diamond in Fig.~\ref{fig:L1L2}b). Since we are seeking a solution with as few nonzero components as possible, this solution is obviously preferable. Note that the other sparse solution $\{x=20/7, y=0\}$ has a larger $\ell_1$ norm of $20/7$ and is therefore not optimal.

A large number of CS algorithms have been developed thanks to very active method development efforts in the CS community.  To clearly illustrate how a sparse solution is obtained, we consider here a conceptually simple iterative algorithm called fixed point continuation (FPC) [E. T. Hale, W. Yin, and Y. Zhang, SIAM Journal on Optimization {\bf 19}, 1107 (2008)] rather than other methods that are more efficient but complicated.
\begin{figure}[htp]
\includegraphics[width =0.45 \linewidth]{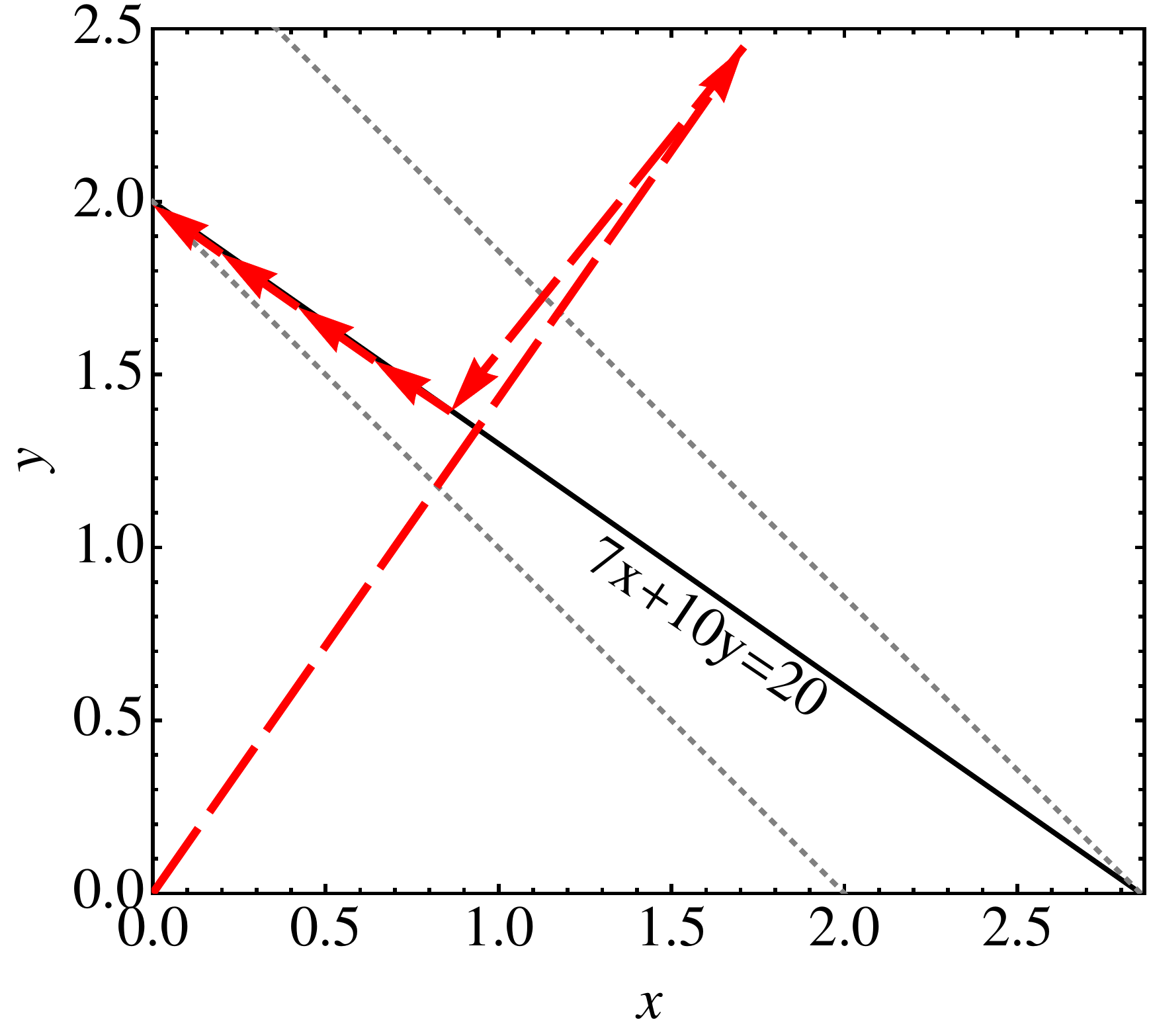}
\caption{Application of the FPC algorithm to the same example. The arrows designate the iterative search process, starting from $(0, 0)$, which optimizes the objective function to find the optimal solution vector $(0, 2)$. For clarity only 7 points along the iterative process are shown. With the first two points, FPC mainly optimizes fitting to the line and moves almost perpendicular to it. The rest of the points show minimization of the $\ell_1$ norm of the solution point by moving towards $\{0, 2\}$, as can be seen with the help of the gray dotted lines representing the $\ell_1$ norm isosurfaces at $| x|+|y| =$ 2 and $20/7$, respectively.}
\label{fig:FPC-example}
\end{figure}
\begin{itemize}
\item The objective function consists of an $\ell_1$ term of the solution vector plus an $\ell_2$ term for the fitting error:
$$ \frac{1}{\mu} \| \mathbf{u} \|_1 + \frac{1}{2} \| \mathbf{f} -  \mathbb{A} \mathbf{ u} \|^2_2$$
\item Input sensing matrix $\mathbb{A}$ ($M \times N $ dimensional), $\mathbf{f}$ ($M$ dimensional), $\mu$
 \begin{enumerate}
    \item Initialize solution vector $\vec{u}^0=\mathbf{0}$,  step size $\tau = \min (1.999,-1.665 M/N + 2.665)$. Normalize $\mathbb{A}$ so that $\alpha_A= \max(\text{eigenvalues}(\mathbb{A}^T \mathbb{A})) \leq 1$. This may easily be accomplished by dividing both $\mathbb{A}$ and $\vec{f}$ by $\sqrt{\alpha_A}$.
    \item for $k= 0, 1, 2, \dots$
     \begin{enumerate}
    \item $\vec{g}^k := \mathbb{A}^T (\mathbb{A} \vec{u}^k - \vec{f})$, where $g$ is the gradient of the $\ell_2$ term, i.e. the direction of steepest descent to minimize the  $\ell_2$ term in the objective function
    \item  $u^{k+1}_n := \text{shrink} \left( u^k_n - \tau g^k_n , \tau/\mu \right)$. In this step, the new solution vector $u^{k+1}$ takes a step in the $-g$ direction and then gets ``shrunk'' to minimize the  $\ell_1$ norm (see below)
    \item  break if converged
     \end{enumerate}
    \item end for
 \end{enumerate}
 \item The shrinkage operator is defined as
$\text{shrink} \left( y, \alpha \right) := \text{sign} (y) \max \left( |y|-\alpha,0 \right)$. It decreases $y$'s absolute value by $\alpha$ and sets $y$ to zero if $|y| \le \alpha$. Shrinkage is a {\bf critical} step in the FPC algorithm to get a sparse solution vector. Parameters that are negligibly small  are truncated to zero in the final solution.
\item Convergence is reached when the gradient $g$ drops below the shrinkage threshold
and the change in the solution vector $u^k$ is sufficiently small.
\end{itemize}

The FPC algorithm is applied to the simple problem $7x+10y=20$ as shown in Fig.~\ref{fig:FPC-example}. The iterative optimization process attempts to minimize both the $\ell_2$ error of fitting ($ \| \mathbf{f} -  \mathbb{A} \mathbf{ u} \|^2_2$) by moving perpendicular towards the solid line (most notably at the first two arrows of Fig.~\ref{fig:FPC-example}) and staying along the solid line, and the $\ell_1$ norm of the solution vector ($ \| \mathbf{u} \|_1 = |x|+|y|$) by shrinking $x$ and $y$. The combined result is that $x$ is shrunk to zero, leaving $y=2$ (Fig.~\ref{fig:FPC-example}).

\subsection{Additional case study results}

\begin{figure}[htp]
\includegraphics[width =0.6 \linewidth]{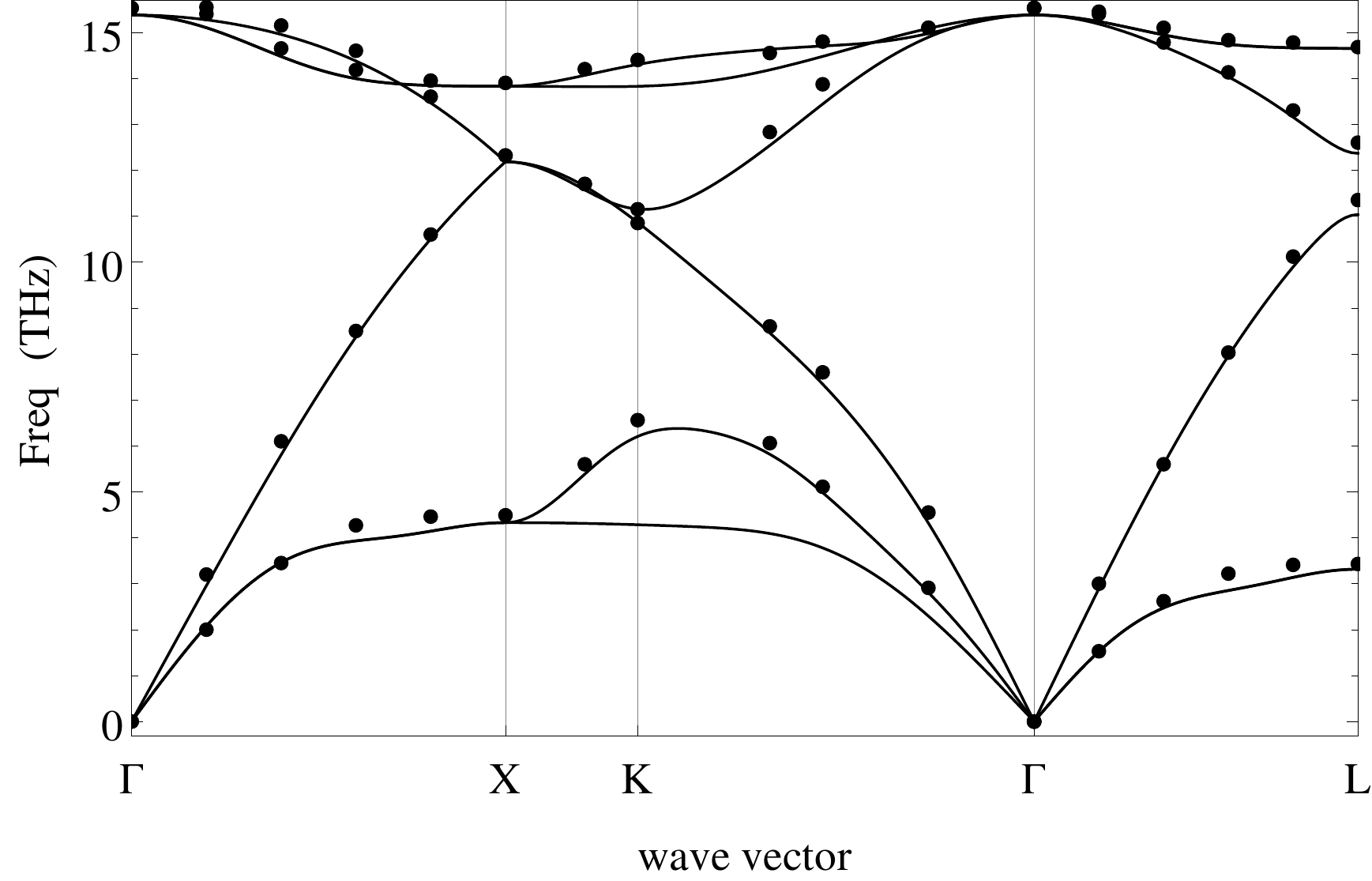}
\caption{Phonon dispersion curve of Si using harmonic force constants extracted by CSLD. The dots are inelastic neutron scattering data from G. Dolling, Inelastic Scattering of Neutrons in Solids and Liquids (IAEA, Vienna, 1963).}
\label{fig:Si-phonon}
\end{figure}

\begin{figure}[htp]
\includegraphics[width =0.6 \linewidth]{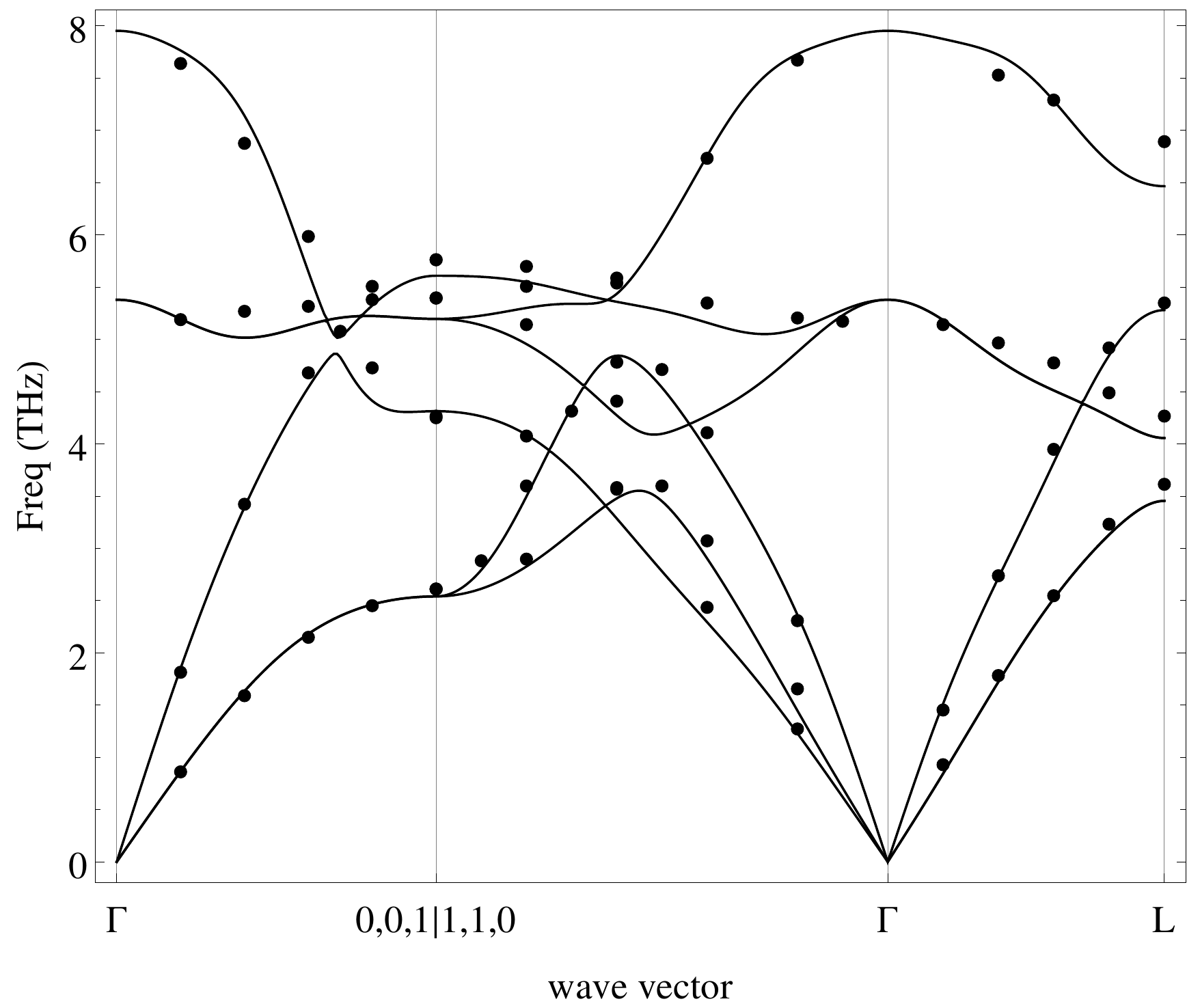}
\caption{The same for NaCl. The dots are inelastic neutron scattering data from G. Raunio, L. Almqvist and R. Stedman,  Phys.\ Rev., {\bf 178}, 1496. (1969). Non-analytical corrections for long-range dipole-dipole interactions have been added using the Parlinski scheme to reproduce the LO-TO splitting. [K. Parlinski, Z. Q. Li, and Y. Kawazoe, Phys. Rev. Lett. 81, 3298 (1998)].}
\label{fig:NaCl-phonon}
\end{figure}

\begin{figure}[htp]
\includegraphics[width =0.6 \linewidth]{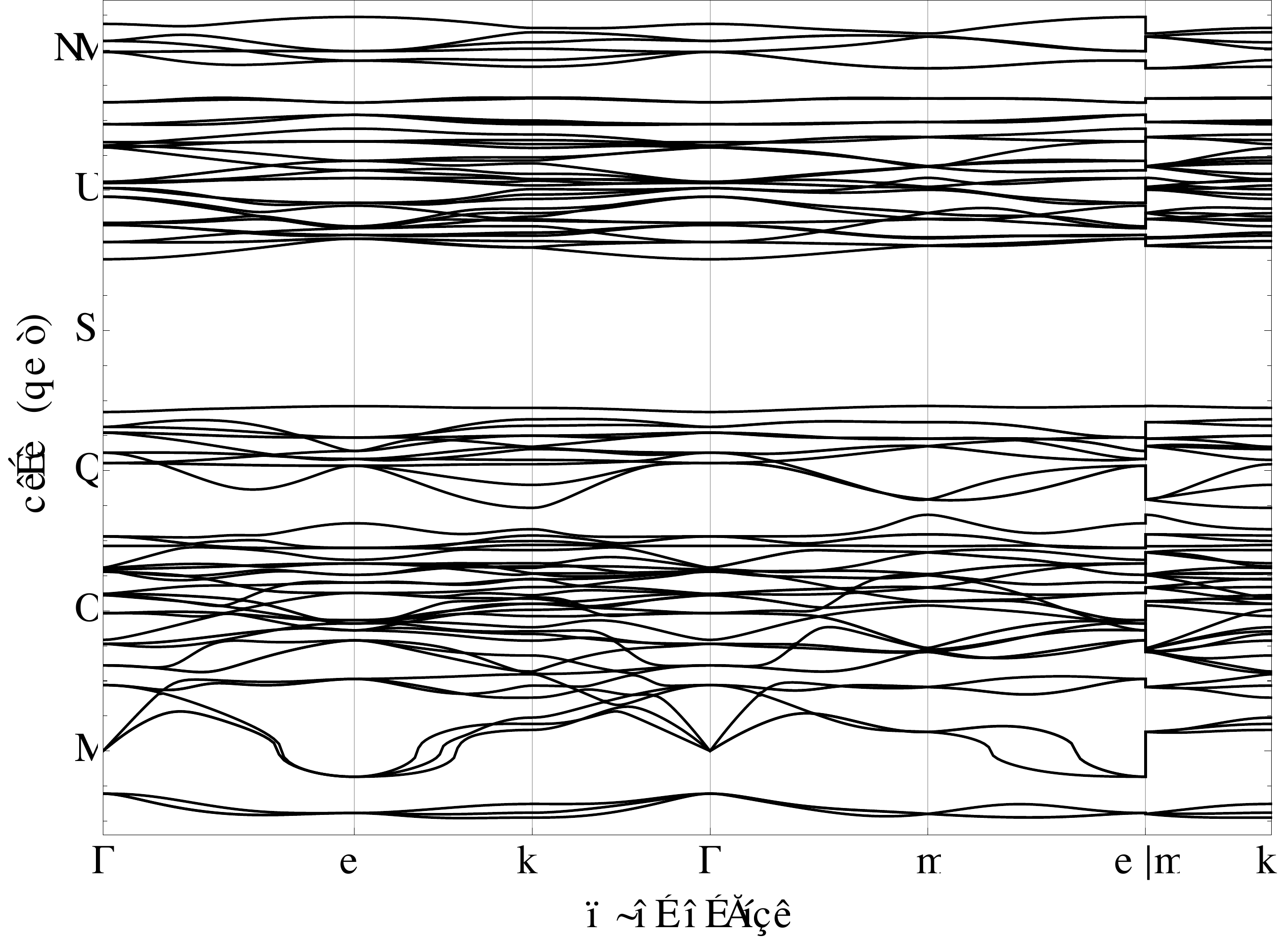}
\caption{The same for Cu$_{12}$Sb$_{4}$S$_{13}$. Since it is a metallic system, non-analytical correction was not necessary. Note that this is calculated with PBE while our previous results were obtained with DFPT using the local density approximation (LDA) [X. Lu, D. T. Morelli, Y. Xia, F. Zhou, V. Ozolins, H. Chi,
X. Zhou, and C. Uher, Adv. Energy Mater. 3, 342 (2012)]}
\label{fig:tetra-phonon}
\end{figure}

\begin{figure}[htp]
\includegraphics[width =0.5 \linewidth]{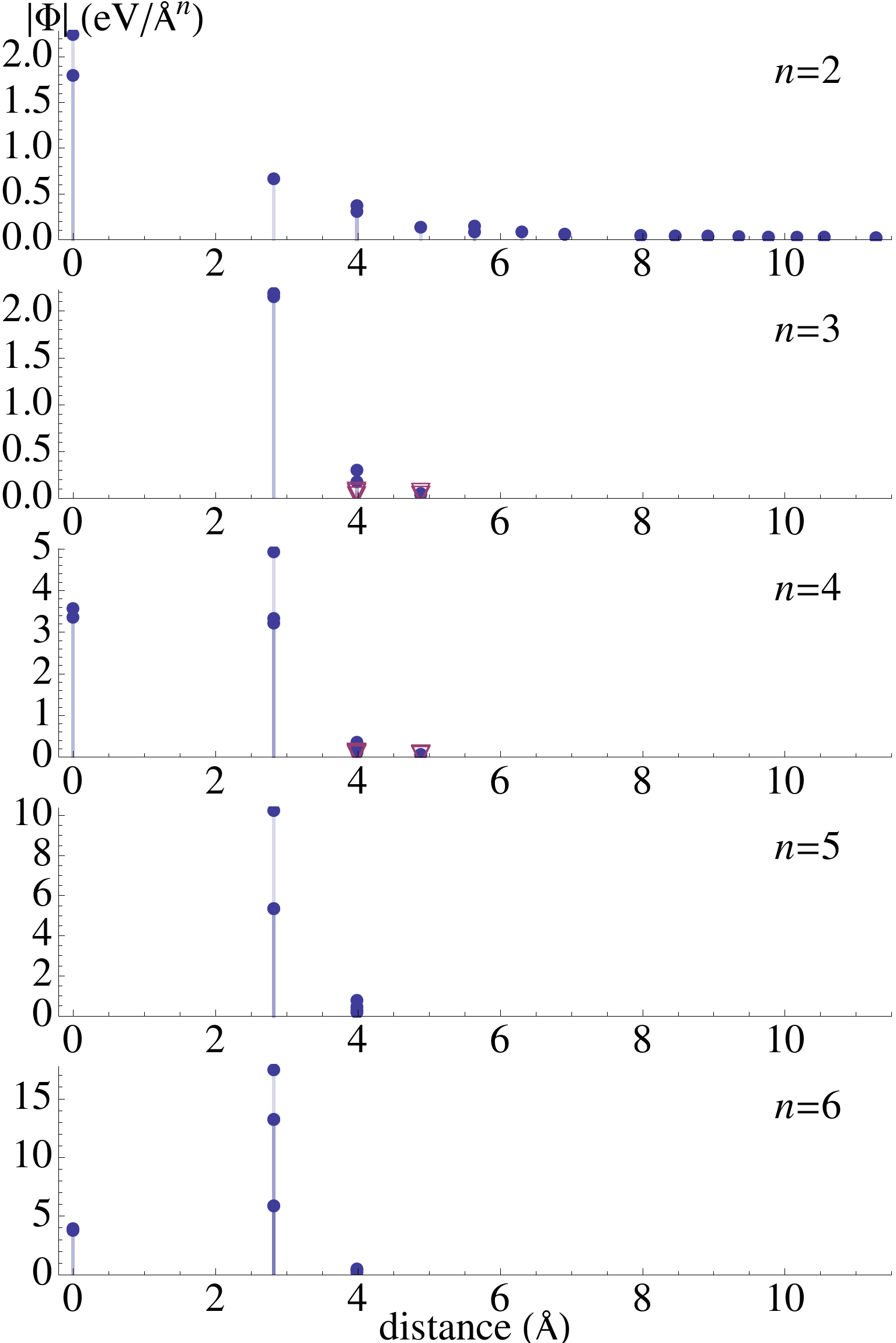}
\caption{Frobenius norm of calculated $n$-th order FCT for NaCl ($n=2$--6) vs.\ interaction distance defined as the maximum distance between interacting atoms. The filled dots represent one (e.g. $\Phi_{aa}$ and $\Phi_{aaaa}$) or two-body (e.g. $\Phi_{ab}$ and $\Phi_{aaabbb}$) FCTs, while the open triangles represent three body interactions, e.g. $\Phi_{abc}$ and $\Phi_{aabc}$. Zero distance implies one-body interactions.  Note the absence of odd-order one-body FCTs $\Phi_{aaa}$ and $\Phi_{aaaaa}$ due to symmetry. Clearly the short range interactions dominate, particularly for higher order anharmonic terms, the most significant of which are nearest-neighbor Na-Cl interactions. Multi-body Na-Cl-Na and Cl-Na-Cl interactions (purple triangles) are found to be very weak.}
\label{fig:NaCl-ECIs}
\end{figure}

\begin{figure}[t]
\includegraphics[width = 0.55 \linewidth]{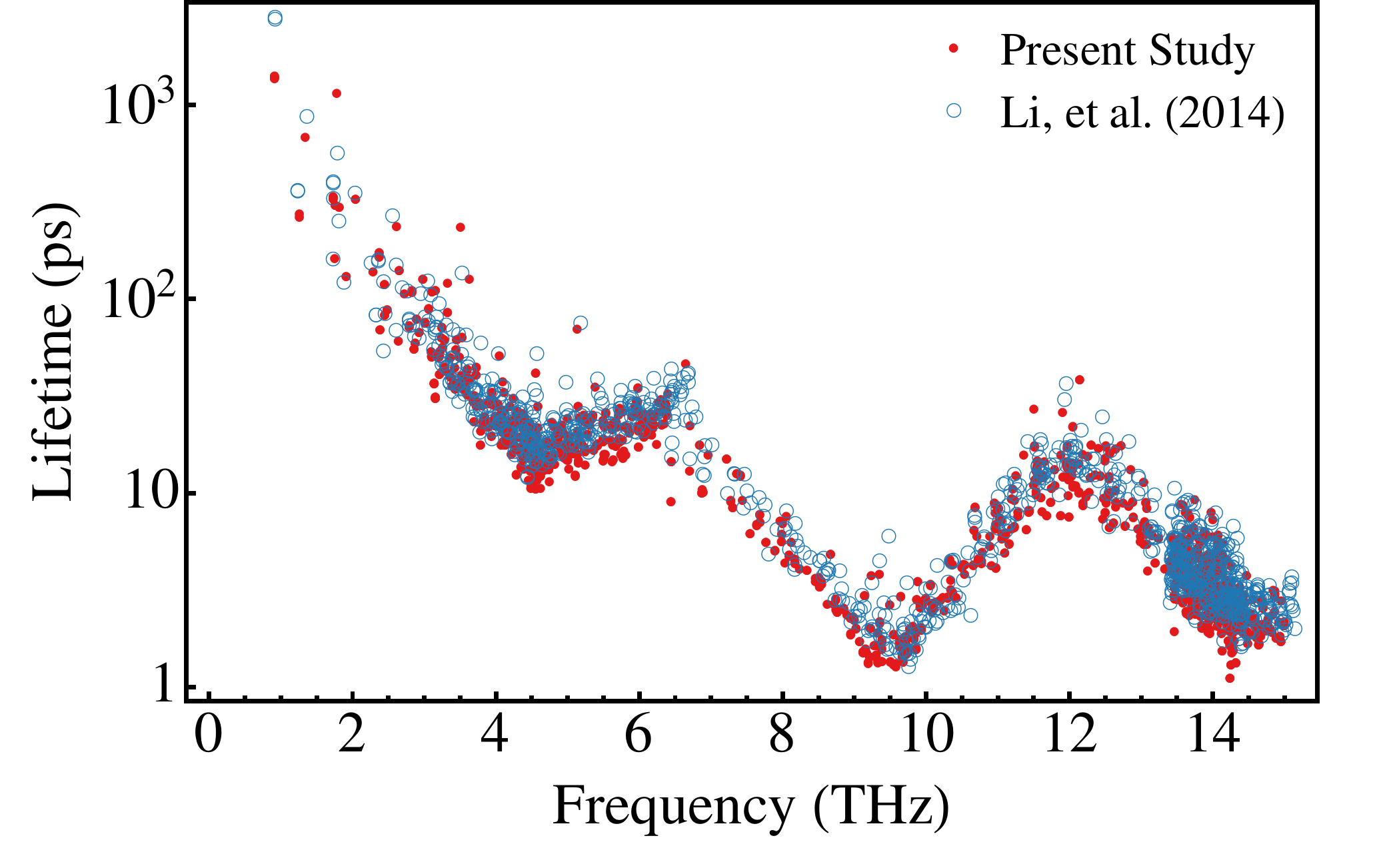}
\caption{Phonon mode lifetimes in silicon at $T=300$~K calculated on a $19 \times 19 \times 19$ mesh of wave vectors. Data from literature were also shown for comparison [W. Li, J. Carrete, N. A. Katcho, and N. Mingo, Comp. Phys. Commun. 185, 1747 (2014)]}
\label{fig:lifetimes}
\end{figure}

\begin{figure}[ht]
    \includegraphics[width =0.55 \linewidth]{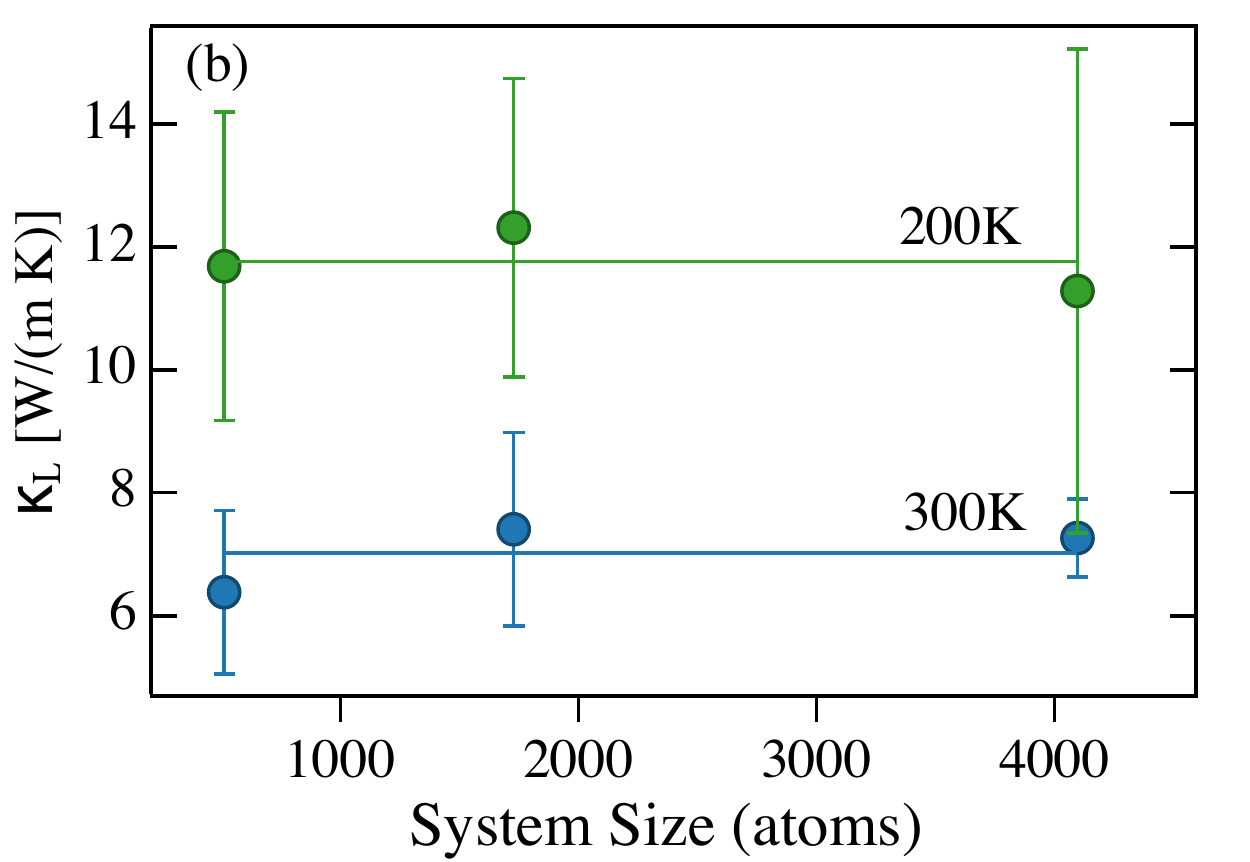}
  \caption{Calculated  $\kappa_{L}$ of NaCl vs.\ the size of the simulation cell. There is no discernible size-dependence in the range of supercells tested.  \label{fig:NaCl-size-dependence}}
\end{figure}


\newpage
\clearpage

\subsection{Performance of LMD}

LMD has been parallelized to take advantage of large computational systems.  The following figures were obtained with sodium chloride system consisting of 1440 atoms. We were not able to perform a single DFT calculation with the computational resources available to us.

\begin{figure}[htp]
\includegraphics[width =0.7 \linewidth]{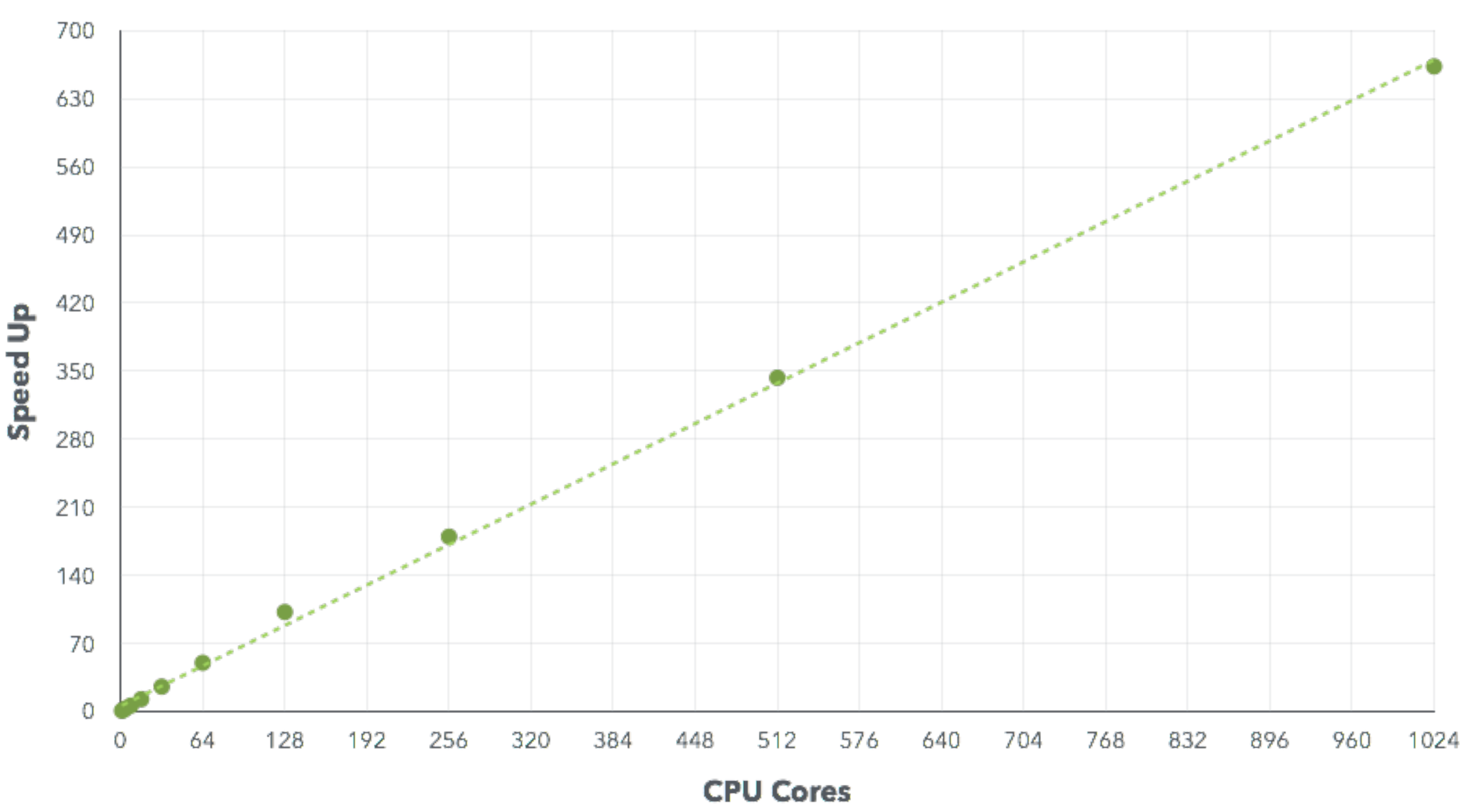}
\caption{LMD exhibits linear performance scaling with increasing parallelization across CPUs.}
\label{fig:LMD-perf}
\end{figure}


\begin{figure}[htp]
\includegraphics[width =0.75 \linewidth]{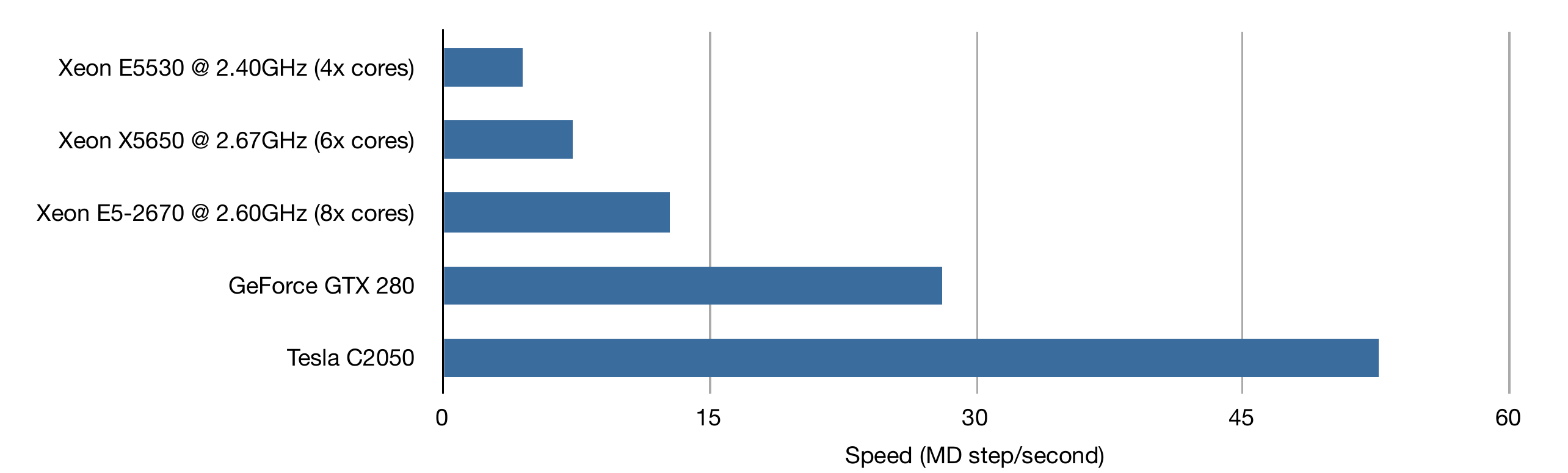}
\caption{LMD performance on a range of different hardware and varying levels of parallelization.  The bottom two devices are graphics processing units (GPUs), both of which outperform conventional CPUs by a large margin.}
\label{fig:LMD-devices_speed}
\end{figure}
\end{document}